# Unraveling the Photophysics of Liquid-Phase Exfoliated Two-Dimensional ReS$_2$ Nanoflakes

**Authors:**

Pieter Schiettecatte,[†,‡,1] Deepika Poonia,[¶,1] Ivo Tanghe,[†,‡] Sourav Maiti,[¶] Michele Failla,[¶] Sachin Kinge,[§] Zeger Hens,[†,‡] Laurens D.A. Siebbeles,[*,¶] and Pieter Geiregat[*,†,‡]

**Affiliation:**

†Physics and Chemistry of Nanostructures, Ghent University, Ghent, Belgium

‡Center for Nano and Biophotonics, Ghent University, Ghent, Belgium

¶Optoelectronic Materials Section, Department of Chemical Engineering, Delft University of Technology, 2629 HZ Delft, The Netherlands

§Materials Research & Development, Toyota Motor Europe, B1930 Zaventem, Belgium

[1] Both authors contributed equally to this work.

**Abstract:**

Few-layered transition metal dichalcogenides (TMDs) are increasingly popular materials for optoelectronics and catalysis. Amongst the various types of TMDs available today, rhenium-chalcogenides ($ReX_2$) stand out due to their remarkable electronic structure, such as the occurrence of anisotropic excitons and potential direct bandgap behavior throughout multi-layered stacks. In this letter, we have analyzed the nature and dynamics of charge carriers in highly crystalline liquid-phase exfoliated $ReS_2$, using a unique combination of optical pump-THz probe and broadband transient absorption spectroscopy. Two distinct time regimes are identified, both of which are dominated by unbound charge carriers despite the high exciton binding energy. In the first time regime the unbound charge carriers cause an increase and a broadening of the exciton absorption band. In the second time regime, a peculiar narrowing of the excitonic absorption profile is observed, which we assign to the presence of built-in fields and/or charged defects. Our results pave the way to analyze spectrally complex transient absorption measurements on layered TMD materials and indicate the potential for $ReS_2$ to produce mobile free charge carriers, a feat relevant for photovoltaic applications.

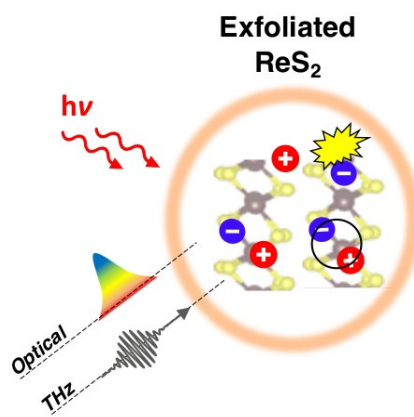

**Introduction**

Over the past few years, layered rhenium chalcogenide compounds have gained significant interest in the scientific community due to their anisotropic electronic and optoelectronic properties[1]. Rhenium disulfide ($ReS_2$) exhibits layer independent optical properties due to electronically and vibrationally decoupled monolayers[2,3]. Among several implications, the possible preservation of a direct electronic band gap for thicker stacks sets $ReS_2$ apart from more conventional transition-metal dichalcogenides (TMDs), which suffer from indirect band gaps for multi-layered stacks[2,4–6]. $ReS_2$ has shown great potential for transistors[7–9], photodetectors[10–12], light-emitting diodes[13], and sensors[14]. Interestingly, unlike conventional two-dimensional (2D) semiconductors, $ReS_2$ has a stable distorted 1T' crystal structure and shows in-plane[15,16] and out-of-plane anisotropy[17,18]. This results in the formation of anisotropic excitons, and as a consequence their fundamental properties such as transport[19], lifetimes, *etc.*, can be selectively modulated by varying the polarization of incident light.

Like other layered semiconductors, $ReS_2$ can be produced either through vacuum epitaxial growth, exfoliation from high-quality crystals, or direct colloidal synthesis[20–23]. Exfoliation, particularly through liquid-phase methods, is a cost-effective means to produce flakes of $ReS_2$ as it combines the advantages of solution processing (low cost, upscaling) while preserving high crystallinity, a persistent issue in direct colloidal synthesis methods[24–26]. An in-depth study of the synthesis of crystalline few-layered $ReS_2$ flakes through liquid-phase exfoliation has been reported recently[27]. Although high-quality $ReS_2$ nanoflakes obtained through such exfoliation procedures are clearly promising materials, the nature and dynamics of elementary optical excitations in these novel materials are not yet known, thereby limiting their potential use in applications such as light emission or photovoltaics.

In this letter, we report the temporal evolution of photogenerated excitons and free charge carriers in highly crystalline liquid phase exfoliated 2D ReS$_2$ nanoflakes with few-monolayer thickness. We photoexcited the sample with optical laser pulses and detected excitons and free charge carriers by time-resolved optical probe and terahertz (THz) conductivity measurements.

The ReS$_2$ thin films studied in this work were produced by drop-casting on quartz from a dispersion containing liquid-phase exfoliated ReS$_2$ flakes in N-methyl-2-pyrrolidone (NMP) and structurally characterized as summarized in Supporting Information S1. Figure 1a (red trace) shows the measured linear optical absorbance spectrum. The dashed black line indicates a fit with a polynomial background and an exciton absorption profile centered at 1.51 eV (820 nm).

The color map of Figure 1b provides a general overview of the transient absorbance (TA) spectroscopy data (see methods section) of the ReS$_2$ film after photoexcitation with a 2.34 eV (530 nm) 180 fs pump pulse and interrogated by a broadband probe spanning the energy range from 1.4 eV (880 nm) up to 2.4 eV (515 nm). The bandgap region shows a complex interplay of spectral effects taking place, a zoom of which is shown in Figure 2a. We observe an interplay of different phenomena, such as exciton profile broadening or shifts, that can super-impose on a negative TA signal ($\Delta A$ <0) due to stimulated emission (SE), or bleach due to state-filling. A description of effects of line shape shifts and broadening on $\Delta A$ spectra is provided in the Supporting Information (S2), but Figure 2b gives a summary of the main effects on a Gaussian transition shape. To analyze the 2D TA map based on these qualitative concepts, we present spectral slices of $\Delta A$ at distinct pump-probe time delays in Figure 2c.

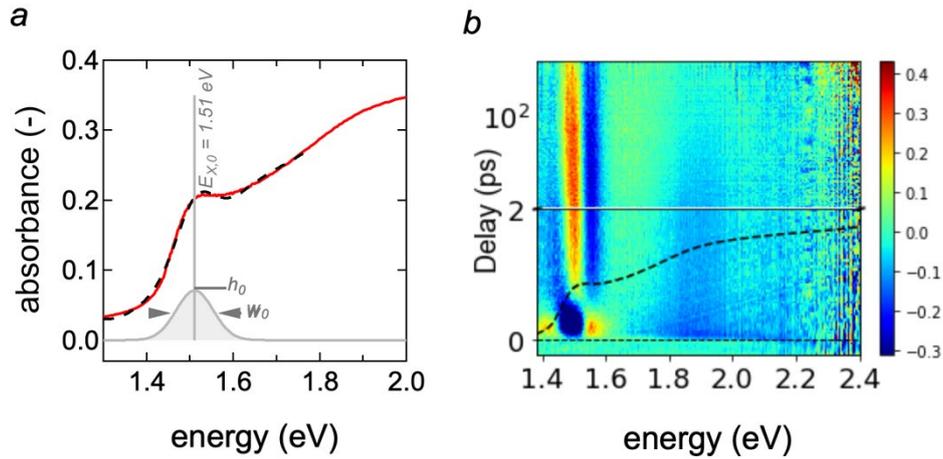

**Figure 1:** *(a) Linear absorption spectrum (red) of a thin film of ReS₂ nanoflakes obtained after dropcasting on quartz. The dashed black line reveals a fit of the absorption spectrum, using a background with an exciton absorption profile centered at 1.51 eV (820 nm, grey shaded area) superimposed. The exciton profile is characterized by a width 'w₀', an amplitude 'h₀' and a central position $E_{X,0}$ (b) Overview of the transient absorbance $\Delta A$ recorded after a film of ReS₂ on quartz was photoexcited with a 180 fs pump pulse with photon energy 2.34 eV and fluence 1.2x10¹³ photons cm⁻² per pump pulse. 2D energy-delay map of $\Delta A$. The black dashed line represents again the linear absorption spectrum.*

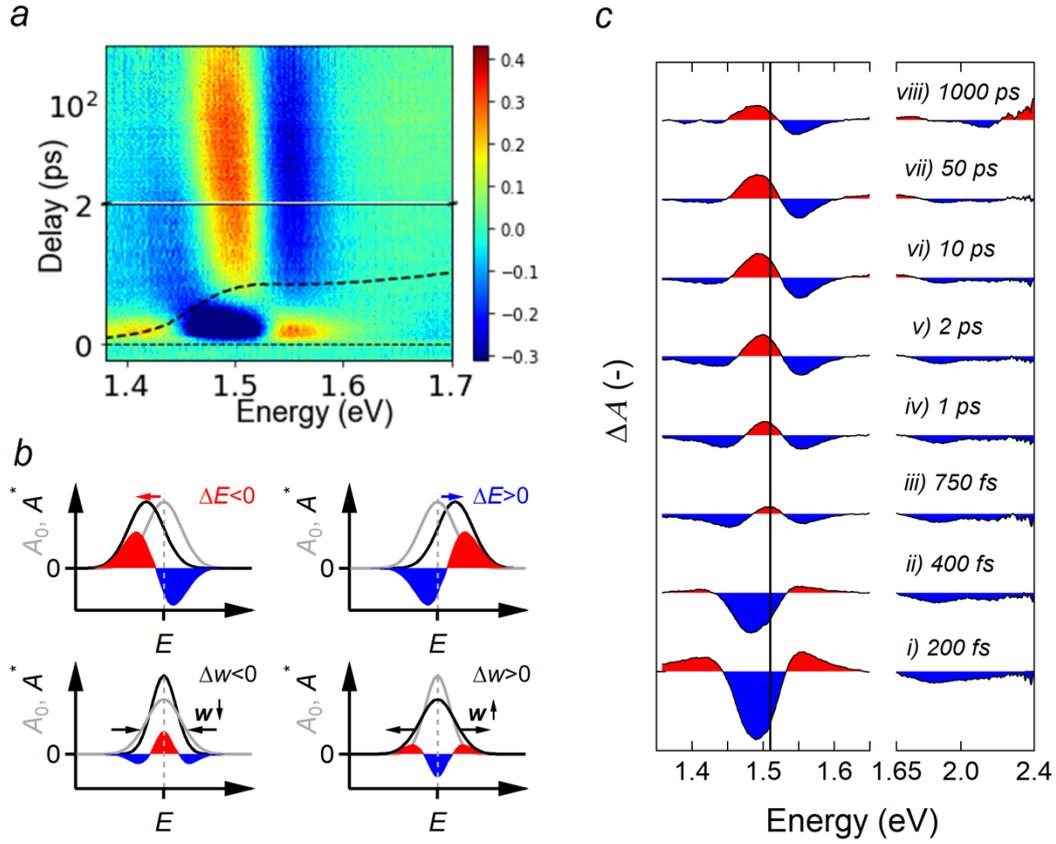

**Figure 2:** *Overview of the transient absorbance $\Delta A$ spectroscopy in the band gap region around 1.51 eV (820 nm) (a) Zoom of the 2D energy-delay map of $\Delta A$ under similar excitation conditions as Figure 1, showing a complex interplay between several spectral effects shown in (b). The black dashed line again represents the linear absorption spectrum. (b) Schematics of pump-induced effects on a Gaussian-shaped optical transition (from top to bottom): spectral redshift, spectral blueshift, linewidth narrowing, and linewidth broadening. (c) Transient absorbance spectra at time delays as indicated, plotted with an offset. For clarity, negative values of $\Delta A$ are filled in blue, positive values of $\Delta A$ in red.*

Immediately after photo-excitation (spectrum i, $t$ = 200 fs), the spectrum exhibits a bleach ($\Delta A < 0$) around the exciton profile sided by two positive photo-induced absorptions (PA, $\Delta A > 0$). We recognize such a second-derivative-like line shape as a combination of SE and/or state-filling induced bleach with a broadening of a Gaussian absorption profile; see Figure 2b and the SI section S2. The bleach minimum is slightly offset towards lower energy, and the PA band appears more intense at the higher energy side. These asymmetries are likely the result of blue-shifted exciton absorption line after the pump pulse, *i.e.,* a shift towards higher energy. At higher probe photon energy (>1.65 eV) , the $\Delta A$ spectrum shows a net negative signal where the ground state absorbance $A_0$ has a relatively uniform slope. Considering the extent of the negative band, it is unlikely to result from state-filling but instead reflects a blueshift of the higher-lying energy states.

Most remarkably, while decaying, the signal at the band edge switches sign over the next few hundreds of femtoseconds (spectra ii-iv) and yields a clear photo-induced absorption sided by two bleach bands at a delay of 1 ps (spectrum iv). Such a feature reflects a narrowed exciton profile after photoexcitation of the sample, *i.e.*, the probe photon produces an exciton with a narrower linewidth than the exciton in the ground state absorption spectrum; see also Figure 2b. At a delay of 750 fs (spectrum iii), the line shape is asymmetric, with the PA maximum slightly offset towards the higher energy side of the exciton profile with a more intense negative tailing at lower energy. The shift to higher energy is super-imposed on the narrowed exciton profile.

At a delay of 1 ps (spectrum iv), the $\Delta A$ feature is fully symmetric, while at pump-probe delays longer than 1 ps (spectra v-vi), the symmetry is reversed, and the initial blueshift has decayed into a redshift; a process that is evidenced by a less intense tailing at lower energies and a shift of the PA maximum to the red side.

At higher energy > 1.65 eV, the blue-shifted continuum absorption has mostly decayed and even yields a slight redshift at decays longer than ≈ 10 ps (spectrum vi), a signature we recognize through the rise of a positive absorption band between 1.6 - 1.7 eV. In contrast, the signature at the band edge decays over a much longer time scale. While decaying (spectra vi-viii), the peak maximum in the $\Delta A$ progressively shifts to the lower energy and obtains sinusoidal-like shape, which is characteristic for a spectral shift of an absorption peak.

In summary, we recognize two dominant regimes in the $\Delta A$ map:

- Regime *I, < 3 ps* : After photo-excitation, the exciton band broadens and blue-shifts. At the same time, we observe a blueshift of the higher-lying energy states.

- Regime *II, > 3 ps - 1 ns* : The initially broadened and blue-shifted exciton transition decays into its mirror image, *i.e.*, it narrows down and shifts to the red. At the same time, the blue-shifted higher-lying states gradually decay.

Having established two distinct regimes of TA dynamics in ReS$_2$, we now turn to providing insights into the nature of photogenerated charges, beit unbound carriers and/or excitons, that are present in each regime and how their presence could induce the observed spectral broadening and shifts in both regimes. The usually very large binding energy of excitons in 2D materials, reported as 117 meV[28] for ReS$_2$, suggest the dominance of tightly bound excitons after energetic relaxation of initially unbound charge carriers produced by the pump laser pulse. However, trapping due to defects and binding energy reduction due to dielectric screening resulting from the presence of free or trapped charge carriers, both result in a shift of the balance from excitons towards charge carriers.

To evaluate the presence and nature of photogenerated free charges versus excitons after photoexcitation, we measured the terahertz (THz) photoconductance after photo-excitation at a pump photon energy of 3.1 eV (400 nm), see methods. Analogous to previous work, we consider the complex THz conductivity signal, $S(t)$, averaged over frequencies in the range of 0.6-1.1 THz, which is given by[29]:

$$S(t) = \phi_{e,h}(t)\mu_{e,h} + \phi_{EX}(t)\mu_{EX} \qquad (1)$$

In the first term of Eq. (1) $\phi_{e,h}(t)$ are the time dependent quantum yields of free mobile charges and $\mu_{e,h}$ is the sum of the complex valued mobility of electrons and holes. The real and imaginary components of $\mu_{e,h}$ are due to motion of charges with velocity in-phase and out-of-phase with respect to the THz field, respectively[30,31]. The second term in Eq. (1) takes into account the contribution of excitons with quantum yield $\phi_{EX}(t)$ and polarizability, $\alpha$, which at a single radian frequency of the THz field ($\omega$) is given by $\alpha = -eIm(\mu_{EX})/\omega$.[29]

The red curve in Figure 3a, due to the real THz conductivity signal from free charges, exhibits an initial fast decay over a time scale of *ca.* 3 ps at an absorbed photon density of $1.1 \times 10^{13}$ photons cm$^{-2}$, which is comparable to the initial variation of the TA spectrum, see Figure 2a-c. Taking the measured initial value of $S_R(t)$ = 10 cm$^2$/Vs (Figure 2a) and assuming the reported electron mobility values for bulk ReS$_2$ in the range 19-34 cm$^2$/Vs[32–34] to be comparable to the hole mobility[32], we estimate the quantum yield of free pairs of electrons and holes to be in the range $\phi_{e,h}(t=0) = 0.20 \pm 0.05$, see Equation 1. After 3 ps, the remaining signal decays on a much longer time scale of several hundreds of picoseconds, possibly due to trapping of remaining charges. Remarkably, the decay of the carrier population is independent of carrier density, as is evidenced by the decay traces of the real THz conductivity in Figure 4a over an order of magnitude in pump fluence. A double-exponential fit to the full decay reveals two lifetimes: $\tau_{fast}$= 0.94 ps and $\tau_{slow}$= 88 ps, see Figure 4b. We use the fast component as a tentative ruler to set the regimes apart, i.e. after 3 decay constants (*ca.* 3 ps) we assume the effects of regime I are finished. Such fluence independent behavior was also observed in flakes of other TMD materials and typically points towards fast trapping of one charge carrier, followed by a non-radiative recombination through a Shockley-Read-Hall mechanism as is shown in Figure 4c, bottom.[24]

The green curve in Figure 3a represents the imaginary component due to the motion of free charges with a velocity component that is out-of-phase with the applied THz field and/or the polarizability of free excitons[29–31]. However, we can show that the exciton polarizability is too small to give a contribution to the THz conductivity signal due to the experimental detection limit (see SI Section S5). Therefore we attribute the observed small imaginary component to backscattering of free charges from randomly oriented stacks of ReS$_2$ flakes and possibly also from localized trap charges. When normalized and plotted on top of each other, the real and imaginary THz conductivity signals follow the same decay kinetics (see SI Figure S9), which strongly corroborates that the real and imaginary component originate from the same species; *i.e.* unbound or 'free' charges. We hence conclude that a significant population of unbound charge carriers are present, in addition to weakly polarizable excitons.

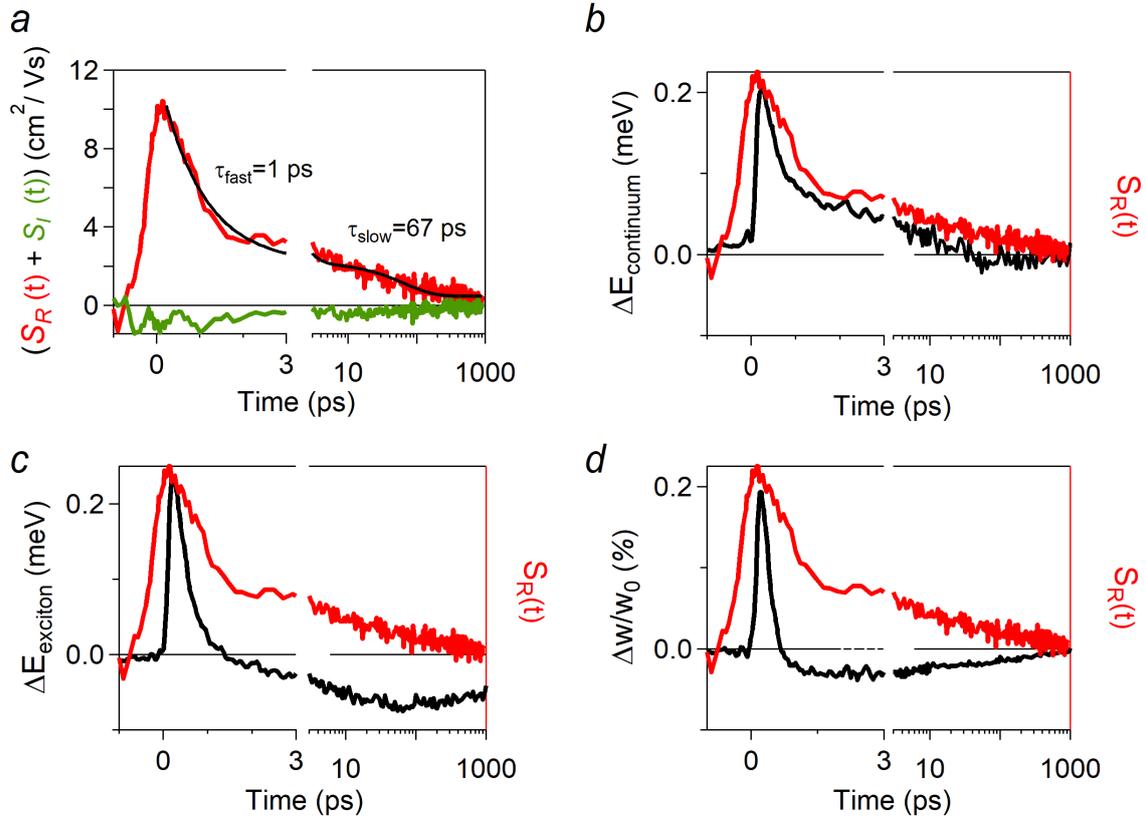

**Figure 3:** *(a) Real and imaginary THz conductivity signal as a function of time obtained after photo-excitation with 3.1 eV energy and at an absorbed photon density, $N_a$ = 1.1x10$^{13}$ cm$^{-2}$. The solid black line is a bi-exponential*

*fit revealing a fast (1 ps) and slow (67 ps) recombination time constant. (b-d) Evolution of the fit parameters (solid black lines) obtained from a fit of a Gaussian G(h, $E_{exciton}$, w) with the constrained area and a background absorbance C(E) to the $\Delta A$ in Figure 1b, including (b) the shift of the background absorbance $\Delta E_{bck}$ and, (c) the energy shift of the exciton $\Delta E_{exciton} = E_X(t) - E_{X,0}$, and (d) the normalized change in the width $\Delta w/w_0$.*

By combining a *qualitative* assessment of the $\Delta A$ maps (Figure 2) and THz spectroscopy (Figure 3a), we have identified two distinct time regimes of free or trapped charge carrier dynamics. To complete this analysis and compare THz to TA dynamics directly, we note that extracting reliable information from the complex and broadband 2D TA map based on slices at fixed probe energies is prone to error. To better quantify the complex interplay of spectral effects - broadening/narrowing, spectral shifts, amplitude decay - that govern the free carrier dominated time response in the two-time regimes, we proceed to fit $\Delta A$ (E, t) to a Gaussian fit function G (h, $E_{exciton}$, w), that accounts for the exciton absorption, and a background absorbance C(E) attributed to higher energy states (see SI S4). Using this procedure, we extract, for each pump-probe time delay the amplitude *h(t)*, the spectral position $E_{exciton}$ (t), and the width *w(t)* of the Gaussian that describes the exciton band, similar as in Figure 1a, and a shift of the background absorbance spectrum $\Delta E_{bck}$. Next, we translate these into differential quantities: $\Delta E_{exciton}$, $\Delta w$, and $\Delta E_{bck}$. The results of this *quantitative* analysis of the $\Delta A$ maps, as shown in Figures 3b-d, reproduce the qualitative assessment put forward earlier, *i.e.,* the complex spectral response is a result of an interplay between line width changes and spectral shifts and is clearly split into two-time regimes. Remarkably, the time evolution of the spectral parameters $\Delta E_{exciton}$, $\Delta w$, and $\Delta E_{bck}$ coincides extremely well with the decay of the real component of the THz conductivity shown in Figure 3a and overlaid in Figures 3(b)-(d).

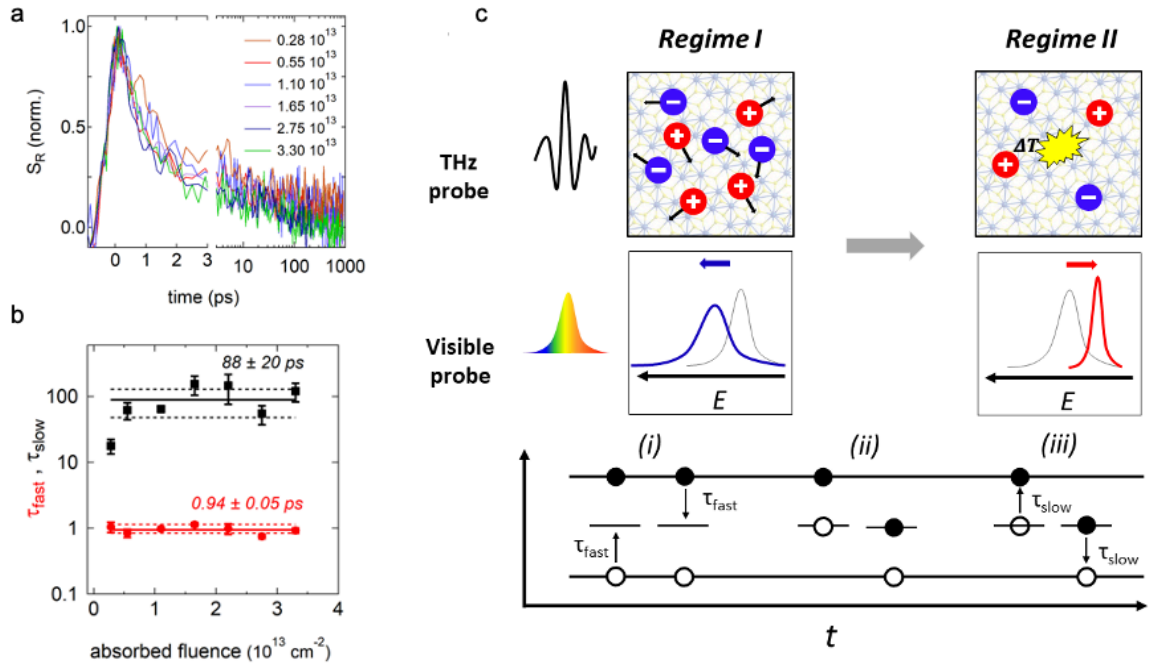

**Figure 4:** (a) Normalized decay traces of the real part of the THz conductivity, a metric for the charge population, for increasing absorbed photon density (0.28 $10^{13}$ – 3.3 $10^{13}$ cm$^{-2}$). (b) Two time constants extracted from a bi-exponential fit to (a) showing density independent recombination dynamics for both fast (red, $\tau_{fast} = 0.94$ ps) and slow (black, $\tau_{slow} = 88$ ps) timescales. *(c) Schematic summary representation of the proposed charge carrier generation and decay in liquid phase exfoliated ReS$_2$, proceeding through two distinct regimes. An optical pump generates predominantly free and mobile charge carriers, as evidenced by THz spectroscopy, that initially broaden and blueshift the exciton profile due to collisional broadening and free carrier screening, respectively. Next, the line broadening and blueshift transition into a peculiar narrowing and redshift on a similar timescale as the THz conductivity decay. The latter indicates a non-radiative recombination event giving rise to the observed spectral signatures, i.e. a redshift due to heating or carrier-capture induced energetic redshifts and a narrowing due to a homogenization of the energy landscape.*

Based on the acquired TA and THz data analysis, we outline two time regimes, both of which are dominated by free mobile charge carriers, which are discussed in detail below, see also Figure 4c:

**Regime *I* ( < 3 ps ): Linewidth broadening and blue-shifts.** The optical pump generates predominantly free charge carriers that initially broaden the exciton resonance due to collisional broadening and blueshift the exciton energy due to a free carrier screening mechanism (Figure 4a). Collisional broadening is responsible for broadening spectral lines in atoms and molecules and absorption and emission lines in semiconductors[35–38]. In addition to the broadening, we also observe a blue shift of the spectrum, which is due to the screening of Coulomb forces resulting in the reduction of the binding energy of excitons[39]. The central position of an exciton resonance $E_X$ depends on the exciton binding energy, $E_b$, and the free-particle bandgap, $E_g$, such that $E_X = E_g - E_b$ [40]. It is well-known that the presence of carriers alters the Coulomb potential experienced by electrons and holes. Screening of attractive electron-hole interactions decreases the exciton binding energy (binding energy reduction, BER), causing blueshifts, while electron-electron (exchange) interactions renormalize the bandgap (bandgap renormalization, BGR), causing red-shifts[41,42]. Both mechanisms work in spectrally opposite directions yet do not fully compensate in general. As such, screening either results in a blue or a redshift of the exciton profile[43]. For instance, blue-shifted exciton transitions as a result of screening have been observed in quasi-2D QWs[36,37,43,44], atomically-thin TMDs[38,39,45], and 2D layered perovskites[46]. In $MoS_2$[38] and $WS_2$[39,45], blueshifts at early delays were always accompanied with a broadening, similar to our observations.

It is worth mentioning that studies typically focus on the lowest-lying exciton state and rarely address the higher-lying energy levels within this context. This makes it more challenging to rationalize a blueshift of the absorption of the higher-lying energy states. Theoretical work addressing the broadband optical response of $MoS_2$ in the presence of excited carriers showed a clear difference between the excitonic response and the higher energy *C* band absorption. While the excitons engaged in a distinctive competitive behavior between the BGR and BER, the *C* peak was relatively stable against increasing carrier densities and even showed signs of a small blueshift[47]. In this regard, Steinhoff *et al.*[47] suggest that an interpretation of higher-lying energy states is far from trivial.

**Regime *II* (> 3ps − 1 ns): Redshifts and linewidth narrowing.** The initial line broadening and blueshift exhibit a cross-over into a peculiar narrowing and redshift on the same timescale as the THz conductivity continues to decrease. Such a transition from a blue to a redshift has been observed in $WS_2$[39,45]. In these studies, the cross-over was attributed to an initial blueshift due to BER, followed by a redshift due to lattice heating (SI S7), arising from non-radiative recombination of electrons and holes on longer timescales (Figure 3b). On the other hand, as outline above, a decay of charge carriers due to recombination would also alter the intricate interplay between BER and BGR. Irrespective of whether the redshift is induced by a temperature effect or by a variation in charge carrier density, tipping the balance between BER and BGR, both are due to the trapping of free mobile charge carriers, an interpretation in agreement with the decay of the real THz conductivity signal.

As carrier-carrier scattering induces a broadening of the line, carrier loss inevitably reduces the width. While such a process accounts for the decay in broadening, it does not explain the small, albeit clearly measurable, narrowing we observe afterward. A line narrowing might occur due to different mechanisms; nonetheless, they again all reflect non-radiative capture events:

1. **Homogenization of the local environment** - A line narrowing can be interpreted as a reduction of the inhomogeneous linewidth of the excitonic optical absorption profile, typically due to a homogenization of the environment. Such homogenization could arise from screening of parasitic in-plane electric fields arising from local potentials, *e.g.* as a result of charged or ionized defects[48,49], or due to pre-existing built-in fields.

We note that such screening of internal electric fields by photogenerated free charges has been studied before by numerous authors on epitaxially grown semiconductor quantum wells[50–53]. The internal electric field in these 2D systems originated from the intrinsic piezoelectric property of the material. Unlike 2H TMDs, such as $MoSe_2$ and $WSe_2$, $ReS_2$ crystallizes in a distorted 1T'-$CdCl_2$ lattice-type, similar to $WTe_2$. Although a $ReS_2$ monolayer (unit cell) is centrosymmetric, exfoliated multi-layered stacks of $ReS_2$ are known to favor an anisotropic AB stacking order where one layer is shifted with respect to the other[9,10]. We, indeed, observed such a stacking order by Raman spectroscopy (SI, S1)[54]. In a similar case, Cobden and co-workers reported the presence of ferroelectricity in bilayer 1T'-$WTe_2$, while it was absent in a monolayer[55]. This distinctive behavior stemmed from the existence of surface dipoles that arise due to the uniquely distorted 1T'-phase. This suggests that an internal field in $ReS_2$ is not unlikely.

Alternatively, Heinz and co-workers outlined that dielectric disorder is the dominant source of inhomogeneities in 2D materials[56] due to the strong effect on the exciton binding energy. Dielectric disorder originates from local fluctuations of the permittivity, for example, due to localized defects. But either by neutralizing internal or parasitic electric fields or by means of improving dielectric order, charge carrier capture into charged defects will always homogenize the energy landscape, thereby reducing inhomogeneous broadening and eventually narrowing the linewidth of the excitonic optical absorption profile.

2. **Interlayer coupling** - An alternative interpretation relates to the interlayer coupling in $ReS_2$. It is well known that exciton profile can be broadened interlayer interactions[17,54]. These effects are subtle, manifesting on changes of the interlayer distance less than 0.1 Angström. A nice example is the existence of two dominant polytypes, labeled as AB (anisotropic) and AA (isotropic). Recently, it has been reported that spectral lines in the AB polytype are broader and blue-shifted with respect to AA. These differences relate to the stronger interlayer coupling between stacked $ReS_2$ layers in the AB polytype – the dominant polytype present in the liquid phase exfoliated $ReS_2$. In this respect, a sudden carrier capture event would locally polarize/charge the sheet and repel adjacent layers, reduces the interlayer coupling – thereby partially undoing differences between AB and AA. As a result, the excitonic absorption profile could narrow.

In summary, we have analyzed the dynamics of photogenerated charge carriers in high quality exfoliated $ReS_2$ nanoflakes using time-resolved pump-probe laser spectroscopy with optical and THz conductivity detection. We identified two distinct time regimes of charge carrier dynamics, both of which are dominated by responses due to unbound charge carriers. In the first time regime, free charge carriers induce collisional broadening of the exciton energy and a blue shift due to screening of the electron-hole attraction. In the second time regime, peculiar line narrowing effects are identified, which could be due to charge carriers reducing effects of built-in fields or charged defects. Besides providing a toolbox to gain insights into a series of complex and interplaying optical nonlinearities in commonly used transient absorption spectroscopy of TMD materials, the generation of unbound and mobile charge carriers in such ultrathin materials is equally interesting from a practical point of view, as mobile charges are highly desired in photovoltaic devices or photodetectors.

## ASSOCIATED CONTENT

**Supporting information** *contains a discussion of the synthesis/structural characterization of ReS$_2$ nanoflakes, a description of lineshape analysis in TA experiments, an in-depth discussion of the spectral deconvolution procedure, calculation of the exciton polarizability and scattering length and a discussion of the lattice heating. Finally, we also report all relevant experimental methods used in this work.*

## AUTHOR INFORMATION

### Corresponding Author

*E-mail: pieter.geiregat@ugent.be

ORCID 0000-0001-7217-8738

### Author contributions

Pieter Schiettecatte and Deepika Poonia contributed equally to this work. Both contributed to the spectroscopy and synthesis, data analysis, and wrote the paper. Ivo Tanghe wrote the analysis code for the spectral deconvolution. Michele Failla assisted with the THz measurements. Sourav Maiti participated in the discussions. Laurens D.A. Siebbeles and Pieter Geiregat supervised the research and wrote the manuscript.

### Acknowledgment

This research received funding from the Netherlands Organisation for Scientific Research (NWO) in the framework of the Materials for sustainability and from the Ministry of Economic Affairs in the framework of the PPP allowance and is also part of the NWO research programme TOP-ECHO with project number 715.016.002. P.S. acknowledges the FWO-Vlaanderen for a fellowship (FWO-SB scholarship, FWO project number: 1S40117N).

## References

(1)   Wang, Q. H.; Kalantar-Zadeh, K.; Kis, A.; Coleman, J. N.; Strano, M. S. Electronics and Optoelectronics of Two-Dimensional Transition Metal Dichalcogenides. *Nat. Nanotechnol.* **2012**, *7* (11), 699–712.

(2)   Tongay, S.; Sahin, H.; Ko, C.; Luce, A.; Fan, W.; Liu, K.; Zhou, J.; Huang, Y.-S.; Ho, C.-H.; Yan, J.; Ogletree, D. F.; Aloni, S.; Ji, J.; Li, S.; Li, J.; Peeters, F. M.; Wu, J. Monolayer Behaviour in Bulk ReS2 Due to Electronic and Vibrational Decoupling. *Nat. Commun.* **2014**, *5* (1).

(3)   Rahman, M.; Davey, K.; Qiao, S.-Z. Advent of 2D Rhenium Disulfide (ReS2): Fundamentals to Applications. *Adv. Funct. Mater.* **2017**, *27* (10), 1606129.

(4)   Dileep, K.; Sahu, R.; Sarkar, S.; Peter, S. C.; Datta, R. Layer Specific Optical Band Gap Measurement at Nanoscale in MoS2 and ReS2 van Der Waals Compounds by High Resolution Electron Energy Loss Spectroscopy. *J. Appl. Phys.* **2016**, *119* (11), 114309.

(5)   Liu, E.; Fu, Y.; Wang, Y.; Feng, Y.; Liu, H.; Wan, X.; Zhou, W.; Wang, B.; Shao, L.; Ho, C.-H.; Huang, Y.-S.; Cao, Z.; Wang, L.; Li, A.; Zeng, J.; Song, F.; Wang, X.; Shi, Y.; Yuan, H.;

Hwang, H. Y.; Cui, Y.; Miao, F.; Xing, D. Integrated Digital Inverters Based on Two-Dimensional Anisotropic ReS2 Field-Effect Transistors. *Nat. Commun.* **2015**, *6* (1).

(6)     Echeverry, J. P.; Gerber, I. C. Theoretical Investigations of the Anisotropic Optical Properties of Distorted 1T ReS2 and ReSe2 Monolayers, Bilayers, and in the Bulk Limit. *Phys. Rev. B* **2018**, *97* (7), 075123.

(7)     Xu, K.; Deng, H.-X.; Wang, Z.; Huang, Y.; Wang, F.; Li, S.-S.; Luo, J.-W.; He, J. Sulfur Vacancy Activated Field Effect Transistors Based on ReS2 Nanosheets. *Nanoscale* **2015**, *7* (38), 15757–15762.

(8)     Liu, F.; Zheng, S.; He, X.; Chaturvedi, A.; He, J.; Chow, W. L.; Mion, T. R.; Wang, X.; Zhou, J.; Fu, Q.; Fan, H. J.; Tay, B. K.; Song, L.; He, R.-H.; Kloc, C.; Ajayan, P. M.; Liu, Z. Highly Sensitive Detection of Polarized Light Using Anisotropic 2D ReS2. *Adv. Funct. Mater.* **2016**, *26* (8), 1169–1177.

(9)     Liao, W.; Wei, W.; Tong, Y.; Chim, W. K.; Zhu, C. Low-Frequency Noise in Layered ReS2 Field Effect Transistors on HfO2 and Its Application for PH Sensing. *ACS Appl. Mater. Interfaces* **2018**, *10* (8), 7248–7255.

(10)    Shim, J.; Oh, A.; Kang, D.-H.; Oh, S.; Jang, S. K.; Jeon, J.; Jeon, M. H.; Kim, M.; Choi, C.; Lee, J.; Lee, S.; Yeom, G. Y.; Song, Y. J.; Park, J.-H. High-Performance 2D Rhenium Disulfide (ReS2) Transistors and Photodetectors by Oxygen Plasma Treatment. *Adv. Mater.* **2016**, *28* (32), 6985–6992.

(11)    Hafeez, M.; Gan, L.; Li, H.; Ma, Y.; Zhai, T. Large-Area Bilayer ReS2 Film/Multilayer ReS2 Flakes Synthesized by Chemical Vapor Deposition for High Performance Photodetectors. *Adv. Funct. Mater.* **2016**, *26* (25), 4551–4560.

(12)   Hafeez, M.; Gan, L.; Li, H.; Ma, Y.; Zhai, T. Chemical Vapor Deposition Synthesis of Ultrathin Hexagonal ReSe2 Flakes for Anisotropic Raman Property and Optoelectronic Application. *Adv. Mater.* **2016**, *28* (37), 8296–8301.

(13)   Wang, J.; Zhou, Y. J.; Xiang, D.; Ng, S. J.; Watanabe, K.; Taniguchi, T.; Eda, G. Polarized Light-Emitting Diodes Based on Anisotropic Excitons in Few-Layer ReS2. *Adv. Mater.* **2020**, *32* (32), 2001890.

(14)   Yang, A.; Gao, J.; Li, B.; Tan, J.; Xiang, Y.; Gupta, T.; Li, L.; Suresh, S.; Idrobo, J. C.; Lu, T.-M.; Rong, M.; Koratkar, N. Humidity Sensing Using Vertically Oriented Arrays of ReS 2 Nanosheets Deposited on an Interdigitated Gold Electrode. *2D Mater.* **2016**, *3* (4), 045012.

(15)   Chenet, D. A.; Aslan, O. B.; Huang, P. Y.; Fan, C.; van der Zande, A. M.; Heinz, T. F.; Hone, J. C. In-Plane Anisotropy in Mono- and Few-Layer ReS2 Probed by Raman Spectroscopy and Scanning Transmission Electron Microscopy. *Nano Lett.* **2015**, *15* (9), 5667–5672.

(16)   Wang, R.; Xu, X.; Yu, Y.; Ran, M.; Zhang, Q.; Li, A.; Zhuge, F.; Li, H.; Gan, L.; Zhai, T. The Mechanism of the Modulation of Electronic Anisotropy in Two-Dimensional ReS2. *Nanoscale* **2020**, *12* (16), 8915–8921.

(17)   Qiao, X.-F.; Wu, J.-B.; Zhou, L.; Qiao, J.; Shi, W.; Chen, T.; Zhang, X.; Zhang, J.; Ji, W.; Tan, P.-H. Polytypism and Unexpected Strong Interlayer Coupling in Two-Dimensional Layered ReS 2. *Nanoscale* **2016**, *8* (15), 8324–8332.

(18)   Hart, L.; Dale, S.; Hoye, S.; Webb, J. L.; Wolverson, D. Rhenium Dichalcogenides: Layered Semiconductors with Two Vertical Orientations. *Nano Lett.* **2016**, *16* (2), 1381–1386.

(19)   Cui, Q.; He, J.; Bellus, M. Z.; Mirzokarimov, M.; Hofmann, T.; Chiu, H.-Y.; Antonik, M.; He, D.; Wang, Y.; Zhao, H. Transient Absorption Measurements on Anisotropic Monolayer ReS. *Small* **2015**, *11* (41), 5565–5571.

(20)    Fujita, T.; Ito, Y.; Tan, Y.; Yamaguchi, H.; Hojo, D.; Hirata, A.; Voiry, D.; Chhowalla, M.;
        Chen, M. Chemically Exfoliated ReS $_2$ Nanosheets. *Nanoscale* **2014**, *6* (21), 12458–
        12462.

(21)    Al-Dulaimi, N.; Lewis, E. A.; Lewis, D. J.; Howell, S. K.; Haigh, S. J.; O'Brien, P. Sequential
        Bottom-up and Top-down Processing for the Synthesis of Transition Metal
        Dichalcogenide Nanosheets: The Case of Rhenium Disulfide (ReS$_2$). *Chem. Commun.*
        **2016**, *52* (50), 7878–7881.

(22)    Keyshar, K.; Gong, Y.; Ye, G.; Brunetto, G.; Zhou, W.; Cole, D. P.; Hackenberg, K.; He, Y.;
        Machado, L.; Kabbani, M.; Hart, A. H. C.; Li, B.; Galvao, D. S.; George, A.; Vajtai, R.;
        Tiwary, C. S.; Ajayan, P. M. Chemical Vapor Deposition of Monolayer Rhenium Disulfide
        (ReS2). *Adv. Mater.* **2015**, *27* (31), 4640–4648.

(23)    Martín-García, B.; Spirito, D.; Bellani, S.; Prato, M.; Romano, V.; Polovitsyn, A.; Brescia,
        R.; Oropesa-Nuñez, R.; Najafi, L.; Ansaldo, A.; D'Angelo, G.; Pellegrini, V.; Krahne, R.;
        Moreels, I.; Bonaccorso, F. Extending the Colloidal Transition Metal Dichalcogenide
        Library to ReS2 Nanosheets for Application in Gas Sensing and Electrocatalysis. *Small*
        **2019**, *15* (52), 1904670.

(24)    Schiettecatte, P.; Geiregat, P.; Hens, Z. Ultrafast Carrier Dynamics in Few-Layer Colloidal
        Molybdenum Disulfide Probed by Broadband Transient Absorption Spectroscopy. *J.
        Phys. Chem. C* **2019**, *123* (16), 10571–10577.

(25)    Zhou, P.; Tanghe, I.; Schiettecatte, P.; van Thourout, D.; Hens, Z.; Geiregat, P. Ultrafast
        Carrier Dynamics in Colloidal WS2 Nanosheets Obtained through a Hot Injection
        Synthesis. *J. Chem. Phys.* **2019**, *151* (16), 164701.

(26)    Coleman, J. N. Liquid Exfoliation of Defect-Free Graphene. *Acc. Chem. Res.* **2013**, *46* (1),
        14–22.

(27) Schiettecatte, P.; Rousaki, A.; Vandenabeele, P.; Geiregat, P.; Hens, Z. Liquid-Phase Exfoliation of Rhenium Disulfide by Solubility Parameter Matching. *Langmuir* **2020**, *36* (51), 15493–15500.

(28) Jadczak, J.; Kutrowska-Girzycka, J.; Smoleński, T.; Kossacki, P.; Huang, Y. S.; Bryja, L. Exciton Binding Energy and Hydrogenic Rydberg Series in Layered ReS2. *Sci. Rep.* **2019**, *9* (1).

(29) Lauth, J.; Failla, M.; Klein, E.; Klinke, C.; Kinge, S.; Siebbeles, L. D. A. Photoexcitation of PbS Nanosheets Leads to Highly Mobile Charge Carriers and Stable Excitons. *Nanoscale* **2019**, *11* (44), 21569–21576.

(30) Ulbricht, R.; Hendry, E.; Shan, J.; Heinz, T. F.; Bonn, M. Carrier Dynamics in Semiconductors Studied with Time-Resolved Terahertz Spectroscopy. *Rev. Mod. Phys.* **2011**, *83* (2), 543–586.

(31) Lloyd-Hughes, J.; Jeon, T.-I. A Review of the Terahertz Conductivity of Bulk and Nano-Materials. *J. Infrared Millim. Terahertz Waves* **2012**, *33* (9), 871–925.

(32) Yu, Z. G.; Cai, Y.; Zhang, Y.-W. Robust Direct Bandgap Characteristics of One- and Two-Dimensional ReS 2. *Sci. Rep.* **2015**, *5* (1), 13783.

(33) Tiong, K. K.; Ho, C. H.; Huang, Y. S. The Electrical Transport Properties of ReS2 and ReSe2 Layered Crystals. *Solid State Commun.* **1999**, *111* (11), 635–640.

(34) Liu, E.; Long, M.; Zeng, J.; Luo, W.; Wang, Y.; Pan, Y.; Zhou, W.; Wang, B.; Hu, W.; Ni, Z.; You, Y.; Zhang, X.; Qin, S.; Shi, Y.; Watanabe, K.; Taniguchi, T.; Yuan, H.; Hwang, H. Y.; Cui, Y.; Miao, F.; Xing, D. High Responsivity Phototransistors Based on Few-Layer ReS2 for Weak Signal Detection. *Adv. Funct. Mater.* **2016**, *26* (12), 1938–1944.

(35) Hindmarsh, W. R.; Petford, A. D.; Smith, G.; Kuhn, H. G. Interpretation of Collision Broadening and Shift in Atomic Spectra. *Proc. R. Soc. Lond. Ser. Math. Phys. Sci.* **1967**, *297* (1449), 296–304.

(36) Leite, R. C. C.; Shah, J.; Gordon, J. P. Effect of Electron-Exciton Collisions on the Free-Exciton Linewidth in Epitaxial GaAs. *Phys. Rev. Lett.* **1969**, *23* (23), 1332–1335.

(37) Schultheis, L.; Kuhl, J.; Honold, A.; Tu, C. W. Ultrafast Phase Relaxation of Excitons via Exciton-Exciton and Exciton-Electron Collisions. *Phys. Rev. Lett.* **1986**, *57* (13), 1635–1638.

(38) Sim, S.; Park, J.; Song, J.-G.; In, C.; Lee, Y.-S.; Kim, H.; Choi, H. Exciton Dynamics in Atomically Thin $MoS_2$ : Interexcitonic Interaction and Broadening Kinetics. *Phys. Rev. B* **2013**, *88* (7).

(39) Cunningham, P. D.; Hanbicki, A. T.; McCreary, K. M.; Jonker, B. T. Photoinduced Bandgap Renormalization and Exciton Binding Energy Reduction in $WS_2$. *ACS Nano* **2017**, *11* (12), 12601–12608.

(40) Wang, G.; Chernikov, A.; Glazov, M. M.; Heinz, T. F.; Marie, X.; Amand, T.; Urbaszek, B. *Colloquium* : Excitons in Atomically Thin Transition Metal Dichalcogenides. *Rev. Mod. Phys.* **2018**, *90* (2).

(41) Schmitt-Rink, S.; Ell, C.; Haug, H. Many-Body Effects in the Absorption, Gain, and Luminescence Spectra of Semiconductor Quantum-Well Structures. *Phys. Rev. B* **1986**, *33* (2), 1183–1189.

(42) Schmitt-Rink, S.; Chemla, D. S.; Miller, D. A. B. Linear and Nonlinear Optical Properties of Semiconductor Quantum Wells. *Adv. Phys.* **1989**, *38* (2), 89–188.

(43)   Wake, D. R.; Yoon, H. W.; Wolfe, J. P.; Morkoç, H. Response of Excitonic Absorption Spectra to Photoexcited Carriers in GaAs Quantum Wells. *Phys. Rev. B* **1992**, *46* (20), 13452–13460.

(44)   Hulin, D.; Mysyrowicz, A.; Antonetti, A.; Migus, A.; Masselink, W. T.; Morkoç, H.; Gibbs, H. M.; Peyghambarian, N. Well-Size Dependence of Exciton Blue Shift in GaAs Multiple-Quantum-Well Structures. *Phys. Rev. B* **1986**, *33* (6), 4389–4391.

(45)   Ruppert, C.; Chernikov, A.; Hill, H. M.; Rigosi, A. F.; Heinz, T. F. The Role of Electronic and Phononic Excitation in the Optical Response of Monolayer WS2 after Ultrafast Excitation. *Nano Lett.* **2017**, *17* (2), 644–651.

(46)   Wu, X.; Trinh, M. T.; Zhu, X.-Y. Excitonic Many-Body Interactions in Two-Dimensional Lead Iodide Perovskite Quantum Wells. *J. Phys. Chem. C* **2015**, *119* (26), 14714–14721.

(47)   Steinhoff, A.; Rösner, M.; Jahnke, F.; Wehling, T. O.; Gies, C. Influence of Excited Carriers on the Optical and Electronic Properties of MoS2. *Nano Lett.* **2014**, *14* (7), 3743–3748.

(48)   Moody, G.; Kavir Dass, C.; Hao, K.; Chen, C.-H.; Li, L.-J.; Singh, A.; Tran, K.; Clark, G.; Xu, X.; Berghäuser, G.; Malic, E.; Knorr, A.; Li, X. Intrinsic Homogeneous Linewidth and Broadening Mechanisms of Excitons in Monolayer Transition Metal Dichalcogenides. *Nat. Commun.* **2015**, *6* (1), 8315.

(49)   STONEHAM, A. M. Shapes of Inhomogeneously Broadened Resonance Lines in Solids. *Rev. Mod. Phys.* **1969**, *41* (1), 82–108.

(50)   Wood, T. H.; Burrus, C. A.; Miller, D. a. B.; Chemla, D. S.; Damen, T. C.; Gossard, A. C.; Wiegmann, W. High-speed Optical Modulation with GaAs/GaAlAs Quantum Wells in a P-i-n Diode Structure. *Appl. Phys. Lett.* **1984**, *44* (1), 16–18.

(51)     Miller, D. A. B.; Chemla, D. S.; Damen, T. C.; Gossard, A. C.; Wiegmann, W.; Wood, T. H.;
         Burrus, C. A. Electric Field Dependence of Optical Absorption near the Band Gap of
         Quantum-Well Structures. *Phys. Rev. B* **1985**, *32* (2), 1043–1060.

(52)     Weiner, J. S.; Miller, D. a. B.; Chemla, D. S.; Damen, T. C.; Burrus, C. A.; Wood, T. H.;
         Gossard, A. C.; Wiegmann, W. Strong Polarization-sensitive Electroabsorption in
         GaAs/AlGaAs Quantum Well Waveguides. *Appl. Phys. Lett.* **1985**, *47* (11), 1148–1150.

(53)     Kuo, Y.-H.; Lee, Y. K.; Ge, Y.; Ren, S.; Roth, J. E.; Kamins, T. I.; Miller, D. A. B.; Harris, J. S.
         Strong Quantum-Confined Stark Effect in Germanium Quantum-Well Structures on
         Silicon. *Nature* **2005**, *437* (7063), 1334–1336.

(54)     Zhou, Y.; Maity, N.; Rai, A.; Juneja, R.; Meng, X.; Roy, A.; Zhang, Y.; Xu, X.; Lin, J.-F.;
         Banerjee, S. K.; Singh, A. K.; Wang, Y. Stacking-Order-Driven Optical Properties and
         Carrier Dynamics in ReS2. *Adv. Mater.* **2020**, *32* (22), 1908311.

(55)     Fei, Z.; Zhao, W.; Palomaki, T. A.; Sun, B.; Miller, M. K.; Zhao, Z.; Yan, J.; Xu, X.; Cobden,
         D. H. Ferroelectric Switching of a Two-Dimensional Metal. *Nature* **2018**, *560* (7718),
         336–339.

(56)     Raja, A.; Waldecker, L.; Zipfel, J.; Cho, Y.; Brem, S.; Ziegler, J. D.; Kulig, M.; Taniguchi, T.;
         Watanabe, K.; Malic, E.; Heinz, T. F.; Berkelbach, T. C.; Chernikov, A. Dielectric Disorder
         in Two-Dimensional Materials. *Nat. Nanotechnol.* **2019**, *14* (9), 832–837.

(57)     Jawaid, A.; Nepal, D.; Park, K.; Jespersen, M.; Qualley, A.; Mirau, P.; Drummy, L. F.; Vaia,
         R. A. Mechanism for Liquid Phase Exfoliation of MoS2. *Chem. Mater.* **2016**, *28* (1), 337–
         348.

(58)     Spoor, F. C. M.; Kunneman, L. T.; Evers, W. H.; Renaud, N.; Grozema, F. C.; Houtepen, A.
         J.; Siebbeles, L. D. A. Hole Cooling Is Much Faster than Electron Cooling in PbSe Quantum
         Dots. *ACS Nano* **2016**, *10* (1), 695–703.

(59) Spoor, F. C. M.; Tomić, S.; Houtepen, A. J.; Siebbeles, L. D. A. Broadband Cooling Spectra of Hot Electrons and Holes in PbSe Quantum Dots. *ACS Nano* **2017**, *11* (6), 6286–6294.

(60) Lauth, J.; Kulkarni, A.; Spoor, F. C. M.; Renaud, N.; Grozema, F. C.; Houtepen, A. J.; Schins, J. M.; Kinge, S.; Siebbeles, L. D. A. Photogeneration and Mobility of Charge Carriers in Atomically Thin Colloidal InSe Nanosheets Probed by Ultrafast Terahertz Spectroscopy. *J. Phys. Chem. Lett.* **2016**, *7* (20), 4191–4196.

(61) Evers, W. H.; Schins, J. M.; Aerts, M.; Kulkarni, A.; Capiod, P.; Berthe, M.; Grandidier, B.; Delerue, C.; van der Zant, H. S. J.; van Overbeek, C.; Peters, J. L.; Vanmaekelbergh, D.; Siebbeles, L. D. A. High Charge Mobility in Two-Dimensional Percolative Networks of PbSe Quantum Dots Connected by Atomic Bonds. *Nat. Commun.* **2015**, *6* (1), 8195.

# Supporting Information

# Unraveling the Photophysics of Liquid-Phase Exfoliated Two-Dimensional ReS$_2$ Nanoflakes

**Authors:**

Pieter Schiettecatte,[†,‡,1] Deepika Poonia,[¶,1] Ivo Tanghe,[†,‡] Sourav Maiti,[¶] Michele Failla,[¶] Sachin Kinge,[§] Zeger Hens,[*,†,‡] Laurens D.A. Siebbeles,[*,¶] and Pieter Geiregat[*,†,‡]

**Affiliation:**

†Physics and Chemistry of Nanostructures, Ghent University, Ghent, Belgium

‡Center for Nano and Biophotonics, Ghent University, Ghent, Belgium

¶Optoelectronic Materials Section, Department of Chemical Engineering, Delft University of Technology, 2629 HZ Delft, The Netherlands

§Materials Research & Development, Toyota Motor Europe, B1930 Zaventem, Belgium

1Both authors contributed equally to the work

## S1: Characterization of Liquid Phase Exfoliated Rhenium Disulfide

**Characterization of Liquid-Phase Exfoliated ReS$_2$**

Rhenium disulfide was exfoliated in *N*-methyl-2-pyrrolidone in accordance with our previous work[1]. The liquid-phase exfoliation (LPE) protocol yields brown, translucent dispersions (Figure S1a) with an excitonic absorption around 820 nm (Figure S1b). Morphological analysis by transmission electron microscopy (TEM, Figure S1c) and atomic force microscopy (AFM, Figures S1d and S1e) reveals that the ReS$_2$ flakes produced by LPE have lateral dimensions of several tens to a hundred nanometer and have an equivalent thickness to ≈ 4±2 ReS$_2$ layers. Elemental analysis by XPS underpins that the flakes have nearly stoichiometric rhenium to the sulphur ratio of 1:1.97-1:1.99 and are oxide-free (Figure S1f). Lastly, the Raman spectrum shows the 18 characteristic vibrations of ReS$_2$ between 100 cm$^{-1}$ and 500 cm$^{-1}$ (Figure S1g), while Re-O vibrations at higher wavenumbers are absent in the spectrum (see inset Figure S1g)[2,3]. A more detailed analysis of the fingerprint region at low wavenumbers reveals that LPE produces the most thermodynamically stable and anisotropically stacked ReS$_2$ polytype, as evidenced by a frequency difference between the A$^1_g$ and the A$^4_g$ modes of ≈ 19-20 cm$^{-1}$ (see Figure S1h)[4,5]. We refer the reader to our previous work for a more elaborate description[1].

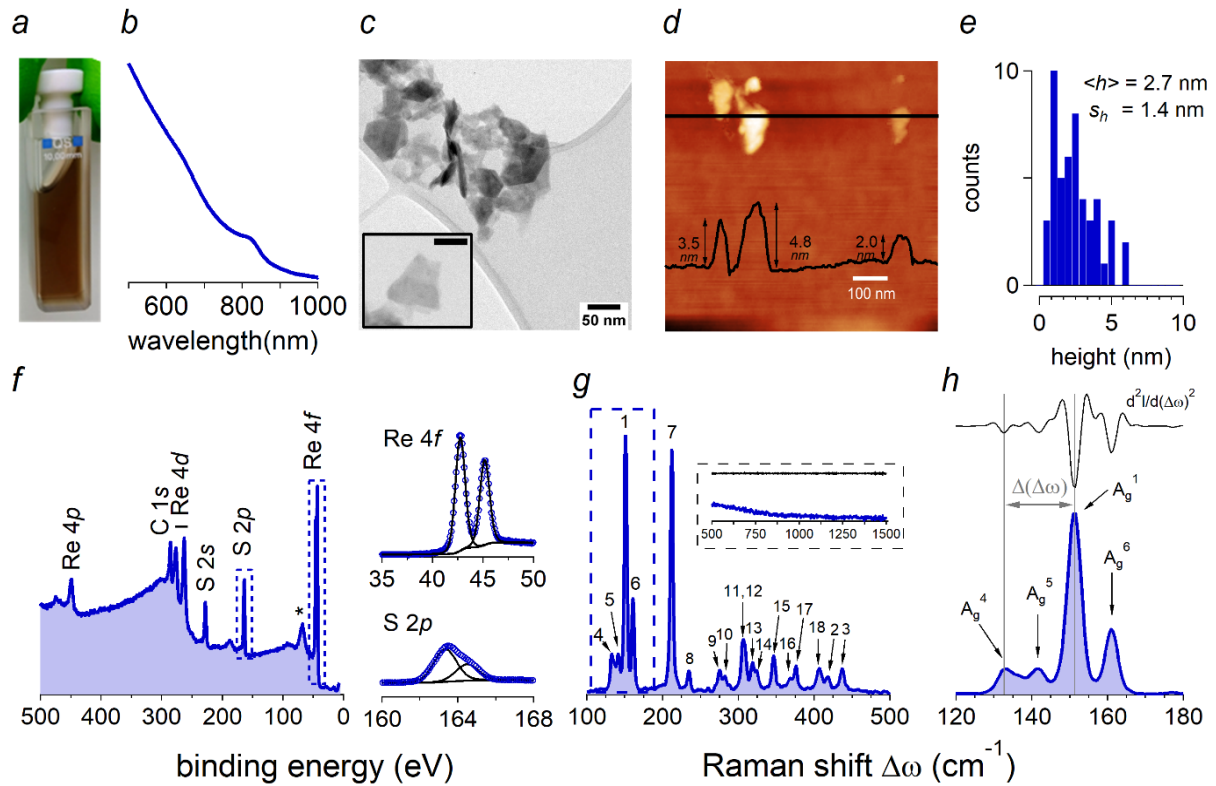

**Figure S1:** *Overview of the material characteristics of liquid-phase exfoliated ReS₂. (a) A photograph of dispersion in N-methyl-2-pyrrolidone. (b) UV-VIS absorbance spectrum. (c) TEM image together with a zoom of an individual flake (scale bar=50 nm). (d) AFM image with a height profile extracted along the black line. (e) AFM height histogram. (f) XPS survey spectrum together with high resolution spectra of the Re 4f and S 2p transitions. The markers represent experimental data points, and the solid line represents a fit of a Lorentzian line shape and a background. (g) Raman spectrum covering the wavenumbers from 100 up to 500 cm⁻¹. We assigned the Raman $A^i_g$ (i=1,...,18) modes in accordance with the reports by the Terrones group[2]. The inset covers the wavenumbers from 500 to 1500 cm⁻¹. (right) A zoom of the Raman spectrum together with the second derivative $d^2I/d(\Delta\omega)$. We marked the frequency difference between the $A^1_g$ and the $A^4_g$ modes is marked as $\Delta(\Delta\omega)$.*

## Optical Absorption of the ReS₂ Film

We describe the absorbance spectrum of the film around the band edge by a Gaussian, accounting for the lowest lying exciton transition, and a polynomial function, describing the background:

$$f(E) = A \exp\left(-\frac{(E-E_0)^2}{w^2}\right) + \sum_{i=0}^{8} C_i E^i \qquad (1)$$

A fit of the optical absorbance (red solid line, Figure S2) to equation 1 (black dashed line, Figure S2) yields an exciton resonance centred around $E_0$ = 1.511 ± 0.001 eV (grey curve, Figure S2).

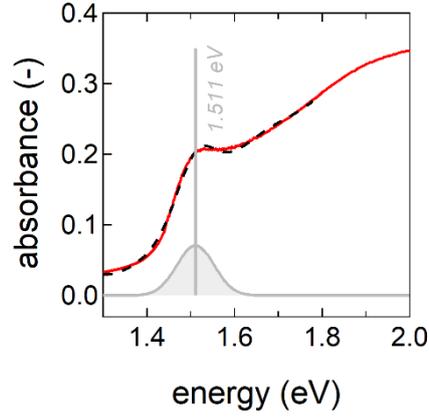

**Figure S2**: *Characterization of the drop-casted ReS$_2$ film on a quartz substrate showing the optical absorption spectrum of the film (red). The black dashed line is a fit to equation 1, yielding a Gaussian-shaped exciton centred around $E_0$= 1.511 ± 0.001 eV (grey).*

We estimate the thickness of the film *t* based on the optical absorbance *A* and the absorption coefficient of rhenium disulphide α:

$$t = \frac{\ln 10\, A}{\alpha} \qquad (2)$$

Using an absorption coefficient α of 8 10$^6$ m$^{-1}$ at 1.5 eV, equation 2 yields a film thickness *t* of 57 nm.

# S2: Understanding Broadening and Spectral Shifts via Derivative Analysis

**Taylor approximations.**

For a transient spectrum $\Delta A$ that is determined by a spectral shift $\delta E$[6] or a change in the linewidth $\Delta\sigma$, we expect a proportionality between the first and second derivative of the linear absorbance $A_0$ to the energy $E$.

$$\Delta A \approx \frac{dA_0}{dE}\delta E \tag{3}$$

$$\Delta A \approx \frac{d^2 A_0}{dE^2}\sigma_0\Delta\sigma \tag{4}$$

We point out that eq 3 and eq 4 are derived by expanding $\Delta A(x + \delta x)$, where $x = E$, $\sigma$ as a Taylor series up to the first order, and thereby only hold for infinitesimal increments $\delta x$ (see mathematical detail below).

**Mathematical detail – Derivation of Taylor Series**

**Proportionality between ΔA and dA$_0$/dE in case of a Spectral Shift δσ**

In the case of a pure spectral shift $\delta E$, the nonlinear absorbance after photoexcitation $A^*$ is a function of $\delta E$, and can be expanded through a Taylor series around the linear absorbance $A_0$:

$$A^* = A_0 + \frac{dA_0}{dE}\delta E + \frac{1}{2}\frac{d^2 A_0}{dE^2}\delta E^2 + \cdots \tag{5}$$

For values of $\delta E$ that are sufficiently small, a Taylor expansion up to the first order yields a direct proportionality between the transient ($\Delta A = A^* - A_0$) and the derivative of the linear absorbance d$A_0$/d$E$:

$$\Delta A \approx \frac{dA_0}{dE}\delta E \tag{6}$$

Accordingly, a value for the spectral shift $\delta E$ can be obtained from the ratio between the transient absorbance and the derivative of the linear absorbance spectrum. Figure S3 illustrates this proportionality between d$A_0$/d$E$ and $\Delta A$ in the case of a pure spectral shift.

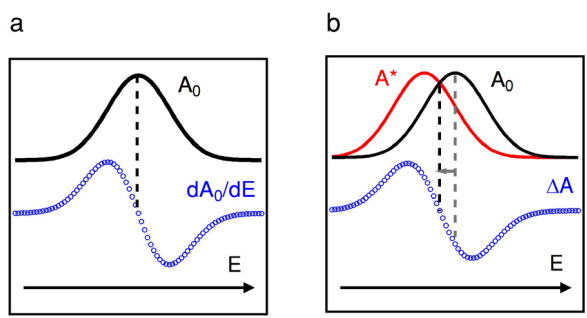

**Figure S3:** *Pictorial representation showing the relationship between a derivative (a) and a spectral shift (b). A spectral shift closely resembles a derivative-like spectrum with the exception that the inflection point is being slightly off-set. The linear absorption $A_0$ is indicated in black, and the excited state absorption $A^*$ in red.*

**Proportionality between $\Delta A$ and d$^2A_0$/d$E^2$ in Case of a Change in Linewidth $\Delta\sigma$**

In the case of a pure spectral broadening or narrowing $\Delta\sigma$, the nonlinear absorbance after photoexcitation $A^*(\sigma)$ can be expanded as a Taylor series around the linear absorbance $A_0(\sigma_0)$:

$$A^*(\sigma_0 + \Delta\sigma) = A_0(\sigma_0) + \frac{dA_0}{d\sigma}\delta\sigma + \frac{1}{2}\frac{d^2A_0}{d\sigma^2}\delta\sigma^2 + \cdots$$

Here again, for small values of $\Delta\sigma$, the transient ($\Delta A = A^* - A_0$) is directly proportional to the derivative of the linear absorbance to the linewidth d$A_0$/d$\sigma$

$$A^*(\sigma_0 + \Delta\sigma) \approx A_0(\sigma_0) + \frac{dA_0}{d\sigma}\delta\sigma \qquad (7)$$

Approximating the absorbance spectrum $A_0$ by a Gaussian function that is characterized by a spectral position $E_0$ and a linewidth $\sigma_0$, the respective derivatives of $A_0$ to $\sigma_0$ and $E$ are given by

$$\frac{dA_0}{d\sigma} = A_0 \frac{E_0^2 - \sigma_0^2}{\sigma_0^3} \tag{8}$$

$$\frac{d^2A_0}{dE^2} = A_0 \frac{E_0^2 - \sigma_0^2}{\sigma_0^4} \tag{9}$$

Combining equations 8 and 9, the equivalence between $\frac{dA_0}{d\sigma}$ and $\frac{d^2A_0}{dE^2}$ reads

$$\frac{dA_0}{d\sigma} = \frac{d^2A_0}{dE^2}\sigma_0 \tag{10}$$

Through equations 7 and 10, the transient $\Delta A$, expressed as the difference between the nonlinear absorbance $A^*(\sigma_0 + \Delta\sigma)$ and the linear absorbance $A_0(\sigma_0)$, now becomes

$$A^*(\sigma_0 + \Delta\sigma) - A_0(\sigma_0) \approx \frac{d^2A_0}{dE^2}\sigma_0\delta\sigma \tag{11}$$

According to Equation 11, a transient absorbance resulting from changes in linewidth has a one-to-one correlation with the second derivative of the linear absorbance spectrum to the energy, and the ratio between both functions yields a value for $\sigma_0\Delta\sigma$. Figure S4 illustrates this proportionality in case of the broadening and the narrowing of a Gaussian.

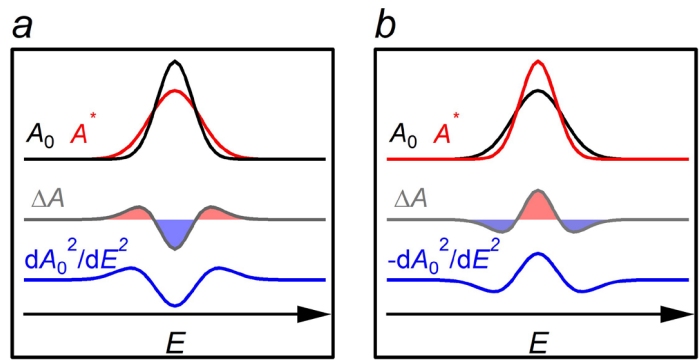

**Figure S4**: *Pictorial representation showing the relationship between (a) the broadening and (b) the narrowing of a Gaussian resonance, the difference spectrum $\Delta A$ and the second derivative of $A_0$ to E. The linear absorbance $A_0$ is indicated in black, the excited state absorbance $A^*$ in red, and the transient absorbance $\Delta A$ in grey.*

# S3: Qualitative Analysis of the Transient $\Delta A$ Spectra based On

## Derivative Analysis

### Regime I – *Linewidth broadening and blueshifts.*

Figures S5a and S5b plot $\Delta A$ after a pump-probe delay of 200 fs (solid red curve) and the averaged $\Delta A$ between $t_0$ and 500 fs (dashed red curve) together with the first derivative $\frac{dA_0}{dE}$ (black, panel a) and second derivative $\frac{d^2A_0}{dE_0}$ (black, panel b). As evident in Figure S5a, the transient has a one-to-one correspondence with $\frac{dA_0}{dE}$ between 1.6-1.9 eV. Notably, the ensuing fixed ratio of $\Delta A$ to the first derivative $\frac{dA_0}{dE}$ is negative and thereby reflects a blueshift of the higher-lying energy states. Using eq. 3, the blueshift $\delta E$, for instance, evaluated at probe energy of 1.75 eV, equals $\approx 0.2$ meV. As shown, later on, this value agrees well with the spectral deconvolution of the transients (see Figure 2 in the main text).

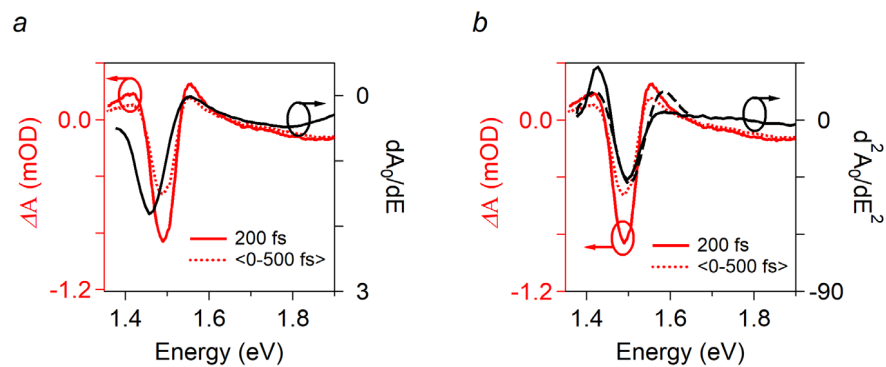

**Figure S5:** *Overview of the derivative analysis in regime I. Comparison of (red) the transient absorbance at a delay of (red, solid line) 200 fs and of (red, dashed line) the averaged signal between $t_0$ and 500 fs to (black, panel a) the first derivative $\frac{dA_0}{dE}$ and to (black, panel b) the second derivative $\frac{d^2A_0}{dE_0}$ to the energy. Take note that the right axis in panel a runs from positive to negative values.*

Around the band-edge, the transient coincides relatively well with minima and maxima in the second derivative $\frac{d^2 A_0}{dE_0}$ to the phonon energy $E$, and the sign of this correspondence marks a broadening of the exciton. For clarity, we plot both the second derivative of $A_0$ (solid black line) and the second derivative of a fitted Gaussian absorption profile to $A_0$ (black dashed line). For an isolated Gaussian, the second derivative is symmetric; that is, a linewidth change has an identical effect at lower and higher energy (black dashed line). In the case of an $A_0$ spectrum consisting of an exciton superimposed on a step-like continuum, a broadening of the exciton line is more pronounced at lower energy (see the solid black line). Irrespective of these considerations, the line symmetry of $\Delta A$ is opposite from both predictions and is consistent with the presence of an additional blueshift. As signified by its correspondence to $-\frac{dA_0}{dE}$, such a blue shift would decrease/increase the intensity at lower/higher energy. Although shifts also contribute to the transient, we can, nonetheless, make a rough estimate on the order of magnitude of the relative broadening $\delta\sigma/\sigma_0$ using eq. 4. In this respect, evaluating $\Delta A$ at different energies, we retrieve values for $\delta\sigma/\sigma_0$ of 0.3 – 0.7%.

**Regime II –** *Linewidth narrowing and redshifts.*

Figures S6a and S6b plot $\Delta A$ after a pump-probe delay of 2 ps (solid red line) and the averaged $\Delta A$ between 1 ps and 1.5 ns (dashed red line) together with the second derivative $\frac{d^2 A_0}{dE_0}$ (black, panel a) and the first derivative $\frac{dA_0}{dE}$ (black line, panel b). At longer time delays, the band edge feature has switched sign and is similar in shape to the mirror image of a second derivative, see Figure S6a. Importantly, such a correlation implies a counter-intuitive narrowing of the exciton band. Similarly, as pointed out above, the shape deviates from a purely narrowed line shape. Given that at those time delays, $\Delta A$ is asymmetric with the lower energy feature being less intense, it also marks a superimposed redshift. As marked by correspondence to $\frac{dA_0}{dE}$, such a redshift would indeed increase $\frac{dA_0}{dE}$ at shorter wavelengths, see Figure S6b.

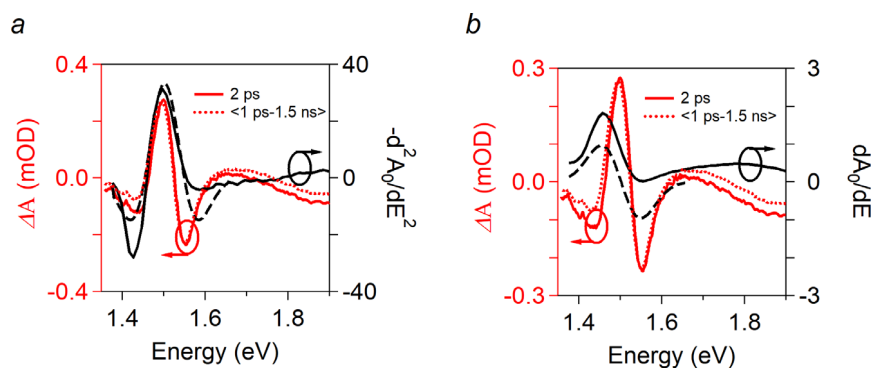

**Figure S6:** *Overview of the derivative analysis in regime II. Comparison of (red) the transient absorbance at a delay of (red, solid line) 2 ps and of (red, dashed line) the averaged signal between 1 ps and 1.5 ns to (black, panel a) the second derivative $-\frac{d^2 A_0}{dE_0}$ and to (black, panel b) the first derivative $\frac{dA_0}{dE}$ to the energy.*

## S4: Spectral Deconvolution of the Transient Absorbance

### Overview of the procedure

A first visual interpretation of the spectra and a comparison of the transients to derivatives of $A_0$ to $E$ marked a complex interplay between shifts and linewidth changes. To better quantify the parameters that govern the transient signal, we describe the ultrafast transient by a Gaussian band-edge exciton $G(E)$ and a background absorbance $C(E)$ that accounts for the higher energy transitions. The Gaussian has (1) a variable central position accounting for spectral shifts, (2) a variable width accounting for line broadening and narrowing effects, and (3) a variable amplitude describing width and areal changes. In addition, the background absorbance spectrum can (4) shift toward lower or higher energy.

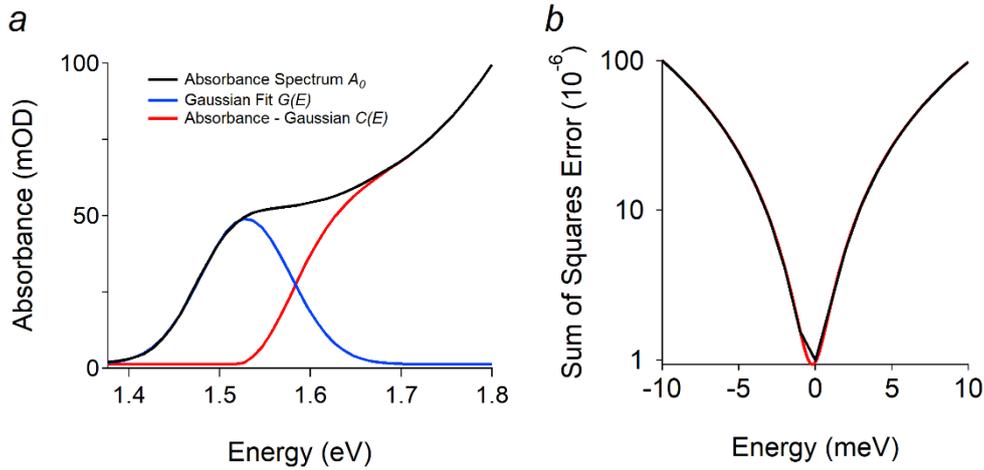

**Figure S7:** *(a) Gaussian fit of the linear absorbance spectrum (black trace), partitioning the Gaussian fit function G(E) (blue trace) and the background absorbance C(E) (red trace). (b) An example of the error analysis.*

In a first step, we describe the linear absorption by a fit to a Gaussian function with the amplitude A, the central position $E_0$, and the width $w$ as fit parameters:

$$G(E) = A.\,exp\,\left(-\frac{(E-E_0)^2}{w^2}\right)$$

Such an approach yields a fitting spectrum as depicted in Figure S7. In this way, the difference between the fitted Gaussian (blue trace) and the linear absorbance spectrum (black trace) yields the background absorbance (red trace).

To fit the transients, we work in two stages. Since a background shift is not a straightforward fitting parameter in numerical software, we move the background in increments $\delta E$. Such an approach does not fit $dA(E,t)$ directly, but rather the difference $dA(E,t) - Lm(E,\ t,\ \delta E)$, with $Lm$ signifying the shifted background function at a background shift $\delta E$. Sweeping the background shift in increments $\delta E$, we scan for the best fit (the fit that provides a minimal error, see Figure S7b), and we extract the other fitting parameters from that particular fit. The to-be fitted function now is

$$dA(E,t) = A_n(t). \, exp \, (-\frac{(E-E_n(t))^2}{w_n(t)^2}) - A. \, exp \, (-\frac{(E-E_0)^2}{w^2})$$

where $A$, $E_0$, and $w$ are fixed to the linear value, and we fit the parameters $A_n(t)$, $E_n(t)$, and $w_n(t)$. This set of fit parameters, together with the background shift $\delta E$, will fit our full model. This procedure is followed for each and every time slice, *i.e.*, at each pump-probe delay.

**Understanding Spectral Shifts and Line Broadening by Spectral Deconvolution of the Transient ΔA Spectra**

Using the aforementioned procedure, we extract, at each pump-probe time delay, the amplitude $h$, the spectral position $E_{exciton}$, and the width $w$ of the Gaussian that describes the exciton band and a shift of the background absorbance $\Delta E_{bck}$. Next, we translate these into differential quantities: $\Delta E_{exciton}$, $\Delta h$, $\Delta w$, and $\Delta E_{bck}$ defined relatively to the respective values before $t_0$, i.e., before photoexcitation.

The primary results of such a spectral deconvolution are summarized in Figure S8 and show the evolution of the parameters mentioned above as a function of the pump-probe delay. In conjunction with a visual interpretation and a derivative analysis, we observe a blueshift of the higher energy states (Figure S8a, $\Delta E_{bck} > 0$), and the values agree with estimates based on eq. 1. Around the band edge, the exciton line is blue-shifted (Figure S8b, $\Delta E_{exciton} > 0$) and broadened (Figure S8c-d, $\Delta w/w_0 > 0$ and $\Delta h/h_0 < 0$). Both results are line with a second-derivative-like line shape that is slightly asymmetric with a more intense PA band at the higher energy side (see Figure 1b, spectrum i).

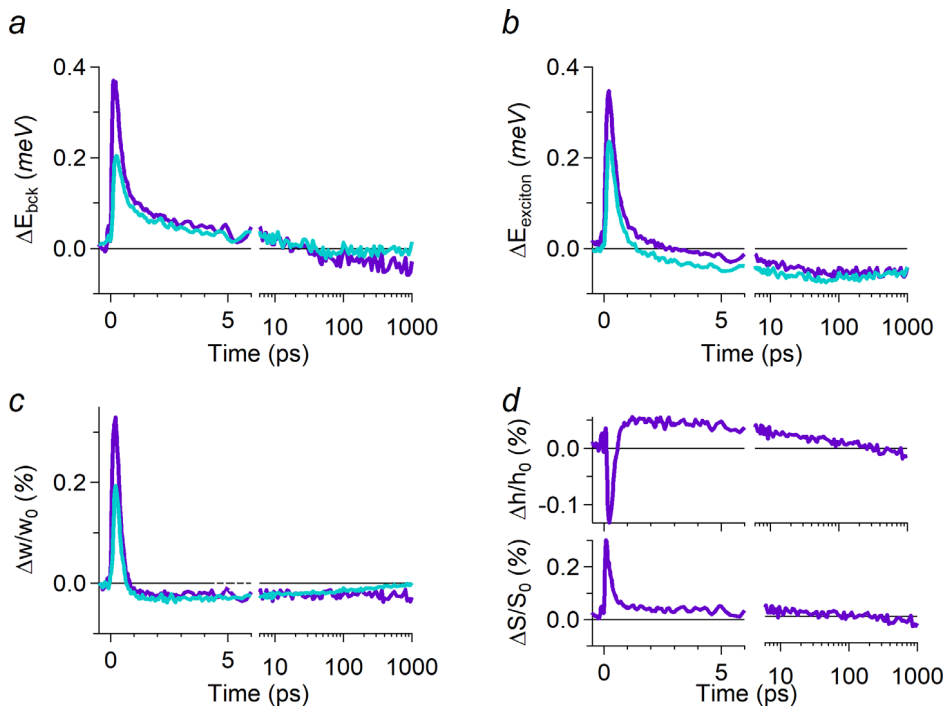

**Figure S8:** *Evolution of the fit parameters that were obtained from a fit to a Gaussian G (h, E$_{exciton}$, w) and a background absorbance C(E) to the ΔA in Figure 1b (cyan and blue curves represent fit with and without constrained area), including (a) the shift of the background absorbance ΔE$_{bck}$ and, (b) the energy shift of the exciton ΔE$_{exciton}$, (c) and (d) the normalized change in the amplitude Δh/h$_0$ of the Gaussian that describes the exciton band.*

Around a time delay of ≈ 1 – 2 ps, the exciton has decayed into its narrowed (Figure S9c- d, $\Delta w/w_0 < 0$ and $\Delta h/h_0 > 0$) and red-shifted (Figure S8b, $\Delta E_{exciton} < 0$) mirror image. Concomitantly, a large fraction of the blue-shifted background (Figure S8a) has decayed. Over time, the exciton redshift reaches a maximum of around ≈ 50 ps, where the background shift starts to fluctuate around zero or becomes even slightly negative. On the other hand, the narrowed exciton line shows little to no decay.

**Limitations -** While the agreement between the quantitative fit and the qualitative analyses presented above is remarkable, we also want to point out a shortcoming of our approach. To account for possible area changes, we implemented the Gaussian amplitude and width as two different fit parameters. In this respect, the area is proportional to the product of the width and the amplitude. As shown in Figure S8d (bottom panel), the Gaussian area seemingly increases after photoexcitation. Likely, this is due to the compensation of over-estimated blueshifts. In this respect, we refitted the data-set while constraining the area of the Gaussian, and we plotted those fit parameters in Figure S8a-c in light blue. Most importantly, the trends we observe are similar. Yet, the initial blueshifts and line broadening are slightly reduced in intensity. In contrast, the exciton redshift gained in intensity at the long delay, and the narrowed line shows a decay over time.

# S5: Exciton polarizability

The exciton polarizability, $\alpha$, is defined as the ratio of the induced dipole moment to the applied electric field and can be theoretically calculated according to $\alpha_{theory} = \frac{2e^2 a_B^2}{E_1 - E_0}$.[7] In this expression, the summation over all higher exciton states is reduced to the first exciton state, and the transition dipole moment is taken equal to the two-dimensional exciton Bohr radius, $a_B$ = 1 nm[8]. Finally, $E_1 - E_0$ is taken equal to exciton binding energy of ReS$_2$, i.e., 117 meV [8]. This yields of value for $\alpha_{theory}$ of $0.27 \times 10^{-35}$ Cm$^2$V$^{-1}$. This number is lower than the experimental upper detection limit. The experimental upper detection limit is calculated through the relation, $\alpha_{exp} = -e\, Im\, (\mu_{EX})/\omega$. Here, $Im\, (\mu_{EX})$ is the experimentally obtained imaginary quantum yield weighted mobility values and $\omega$ is THz probe frequency. This results in $\alpha_{exp}$ =2.7 $\times 10^{-35}$ Cm$^2$V$^{-1}$.

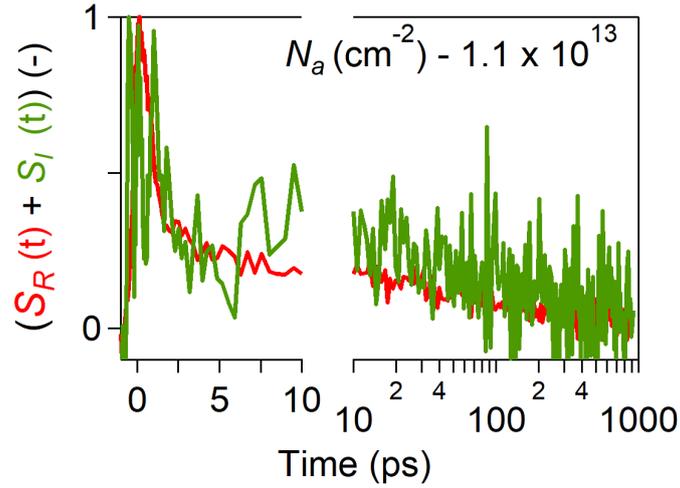

**Figure S9**: *Normalized real and imaginary components of the THz conductivity showing similar relaxation dynamics. This is interpreted as both components resulting from free mobile charge carriers.*

## S6: Charge Carrier Scattering Length

The free charge carriers scattering length can be calculated using, $L = \sqrt{\frac{4\mu_{dc}k_B T t_{osc}}{e}}$. Here, $\mu_{dc}$ is the dc mobility of free charges, $t_{osc}$ is the time period of a single THz pulse (1 $ps$), $k_B$ is the Boltzmann constant, $T$ is taken at room temperature, and $e$ is the electronic charge. From literature, the obtained value of $\mu_{dc}$ for 4.5 nm thick layer of ReS$_2$ is 30 cm$^2$/Vs [9]. Thus, the scattering length of charges during a period of ($t_{osc}$) the THz field is 17 nm, which is higher than the flakes thickness.

## S7: Lattice Heating

The redshift of the exciton peak position on longer time scales has been measured as a function of absorbed pump fluence. Figure S10 summarizes the main results of this analysis by plotting the evolution of ΔA, evaluated at a pump-probe delay of 1 ns, as a function of the absorbed pump fluences. At low fluences (Figure S10, i-ii), the band edge exciton is clearly narrowed due to the proportionality of ΔA to the second derivative of A$_0$ to E (see Figure S4b and Figure S10c). However, when progressively increasing the carrier density fluences (Figure S10, iii-iv), the transient redshifts. We can recognize this patterns due to its close resemblance with the first derivative of $A_0$ to E to the energy E (see Figure S3 and Figure S10b). The effect of absorbed photon fluence on the redshift of the exciton peak position supports the assignment of an increased lattice temperature.

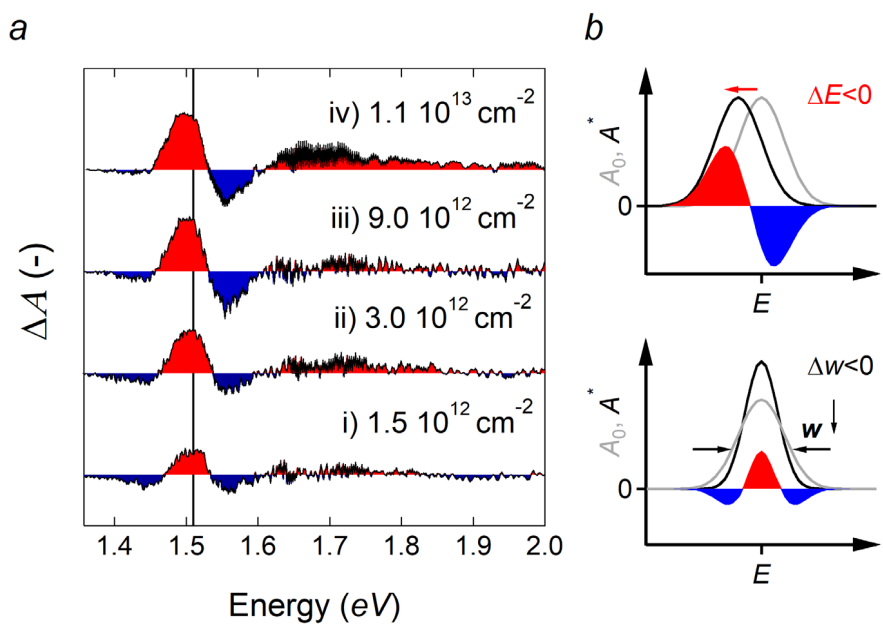

**Figure S10**: *The redshift of the exciton peak on increasing absorbed pump fluence at a probe delay time of 1 ns.*

# S8: Methods

## Structural Characterization

For transmission electron microscopy (TEM), we diluted the dispersions of ReS₂ in isopropanol and cast a drop on a Holy Carbon-Cu grid (200 mesh, 50 microns). We removed traces of residual volatile components under a reduced atmosphere by drying samples in a vacuum tube connected to a Schlenk line or in the antechamber of a nitrogen-filled glove box. Low-resolution TEM images were taken on a Cs-corrected JEOL 2200-FS TEM at an acceleration voltage of 200 kV. For atomic force microscopy (AFM), we prepared films by casting a diluted dispersion in isopropanol on a preheated silica wafer and imaged the films on a Bruker Dimension Edge equipped with a NCHV probe using tapping mode in air. Particle size distributions were measured by Dynamic Light Scattering (DLS) on a Malvern Nano ZS Zetasizer equipped with a red He-Ne laser ($\lambda$=633 nm). For X-ray photoelectron spectroscopy (XPS) and Raman spectroscopy, we casted a concentrated dispersion of $ReS_2$ in NMP on a microscopy glass slide and removed the solvent under vacuum in the antechamber of a glove box. XPS spectra were recorded on an S-Probe Monochromatized XPS spectrometer using a monochromatized Al K$\alpha$ source operated at 1486 eV as a probe. Raman measurements were conducted on a Bruker Optics Senterra dispersive Raman spectrometer attached to an Olympus BX51 microscope and coupled to a thermoelectrically cooled CCD detector, operating at $-65°$C. The spectrometer has an XYZ motorized (automatic) stage for positioning and focusing. Spectra were obtained in the spectral range of 60 to 1560 cm$^{-1}$ and with a spectral resolution of 3 − 5 cm$^{-1}$ using a green Nd:YAG (532 nm) laser. The experimental conditions were set at 5 accumulations of 10 seconds, and sufficiently low laser power (0.1 mW) was used. All the samples were measured under a ×50 magnification objective (numerical aperture (NA) of 0.75) with a spot size of 4 $\mu$m.

### Synthesis and film preparation:

**D**ispersions of $ReS_2$ were produced by liquid phase exfoliation in *N*-methyl-2-pyrrolidone as reported by us previously[27]. In brief, we mixed 50 mg $ReS_2$ bulk powder ($\geq$ 99%, Alfa Aesar,89482.04) together with 25 mL *N*-methyl-2-pyrrolidone (NMP, $\geq$ 99.0 %, Merck) in a centrifugation tube . We sonicated the resulting mixture for a total duration of 6 hours in a sonication bath (USC-THD, VWR) while keeping the bath temperature below 40°C to avoid auto oxidation of NMP[57]. Subsequently, we subjected the dispersion to centrifugation (Eppendorf model 5804, 30 minutes, 1700 rpm) to separate exfoliated flakes from unexfoliated material. The experimental parameters were optimized in our previous work. The supernatant that contains exfoliated $ReS_2$ flakes was decanted from the sediment and stored for further use. Such dispersions of $ReS_2$ in *N*-methyl-2-pyrrolidone were colloidally stable for months with limited sedimentation over time. Finally, a $ReS_2$ film was prepared by drop-casting a $ReS_2$ dispersion on a quartz substrate, followed by annealing at 30°C for 20 minutes.

**Transient absorption (TA)**

TA measurements were performed by optically exciting and probing the sample with ultrashort laser pulses, analogous to previous studies[58–60]. The TA signal was detected using broadband probe pulses generated in a sapphire (500-1600 nm) crystal. A Yb: KGW oscillator (Light Conversion, Pharos SP) is used to produce 180 fs pulses with a 1028 nm wavelength, at a 5 kHz frequency. The pump is obtained by sending the fundamental beam through an Optical Parametric Amplifier (OPA) equipped with a second harmonic module (Light conversion, Orpheus) to perform nonlinear mixing and achieve wavelengths of 310-1330 nm. Pump and probe pulses overlap on the sample at a relatively small angle (~$8^0$), after which the pump pulses are dumped, and the probe light is collected at a detector. The pump-induced change in absorbance is calculated using $\Delta A = \log_{10}(I_{off} / I_{on})$, where I is the incident light on the detector with either pump on or pump off. TA data is further corrected for probe-chirp using a polynomial function. In present measurements, differential reflectance is not considered, and hence, differential transmittance is considered equal to differential absorbance. The pump photon fluence was estimated by measuring with a thermopile sensor (Coherent, PS19Q).

**Optical pump-THz probe measurements**

The terahertz conductivity setup is based on a laser system with Mira Oscillator and a Legend HE-USP regenerative amplifier (by Coherent Inc). The measurements were performed by photoexciting the sample by pump pulses of 60 fs duration and probing photogenerated charge carriers and/or excitons using single-cycle THz pulses which are generated in a nonlinear zinc telluride (ZnTe) crystal via optical rectification[29,61]. The detection of the THz wave form takes place in a ZnTe crystal by spatially overlapping single-cycle THz pulses with a chirped optical laser pulse centered at 800 nm, such that the entire THz waveform is detected by a single laser shot. The photogeneration quantum yield of charge carriers and excitons and their decay kinetics are obtained from the difference, $\Delta E(t_p, t) = E_0(t_p) - E_{excited}(t_p, t)$. Here, $E_{excited}(t_p, t)$ and $E_0(t_p)$ are the transmitted THz pulses after and before photoexcitation. The time $t_p$ is the detection time of the THz probe-pulse and $t$ is the time delay between the THz probe and the laser pump pulse used to photoexcited the sample. The quantum yield weighted real and imaginary mobility of free charges and excitons is related to the differential THz signal $\Delta E(\nu, t)$, according to $S(\nu, t) = \dfrac{(1 + n_s) c \varepsilon_0}{e N_a} \dfrac{\Delta E(\nu, t)}{E_0(\nu, t)}$. Here, $\Delta E(\nu, t) = E_0(\nu) - E_{excited}(\nu, t)$, where $E_0(\nu)$ and $E_{excited}(\nu, t)$ are obtained after Fourier transformation of $E_0(t_p)$ and $E_{excited}(t_p, t)$, while $n_s$, $c$, $\varepsilon_0$ and $e$ are the refractive index of the quartz substrate ($\sqrt{\varepsilon_s} = n_s = 2$) speed of light in a vacuum, vacuum permittivity, and elementary charge, respectively. $N_a$ is the absorbed pump laser photon density per unit area.

# References

(1)    Schiettecatte, P.; Rousaki, A.; Vandenabeele, P.; Geiregat, P.; Hens, Z. Liquid-Phase Exfoliation of Rhenium Disulfide by Solubility Parameter Matching. *Langmuir* **2020**, *36* (51), 15493–15500.

(2)    Vuurman, M. A.; Stufkens, D. J.; Oskam, A.; Wachs, I. E. Structural Determination of Surface Rhenium Oxide on Various Oxide Supports (Al2O3, ZrO2, TiO2 and SiO2). *J. Mol. Catal.* **1992**, *76* (1), 263–285.

(3)    McCreary, A.; Simpson, J. R.; Wang, Y.; Rhodes, D.; Fujisawa, K.; Balicas, L.; Dubey, M.; Crespi, V. H.; Terrones, M.; Hight Walker, A. R. Intricate Resonant Raman Response in Anisotropic ReS2. *Nano Lett.* **2017**, *17* (10), 5897–5907.

(4)    Urban, J. M.; Baranowski, M.; Kuc, A.; Kłopotowski, Ł.; Surrente, A.; Ma, Y.; Włodarczyk, D.; Suchocki, A.; Ovchinnikov, D.; Heine, T.; Maude, D. K.; Kis, A.; Plochocka, P. Non Equilibrium Anisotropic Excitons in Atomically Thin ReS $_2$. *2D Mater.* **2018**, *6* (1), 015012.

(5)    Qiao, X.-F.; Wu, J.-B.; Zhou, L.; Qiao, J.; Shi, W.; Chen, T.; Zhang, X.; Zhang, J.; Ji, W.; Tan, P.-H. Polytypism and Unexpected Strong Interlayer Coupling in Two-Dimensional Layered ReS2. *Nanoscale* **2016**, *8* (15), 8324–8332.

(6)    Geiregat, P.; Houtepen, A.; Justo, Y.; Grozema, F. C.; Van Thourhout, D.; Hens, Z. Coulomb Shifts upon Exciton Addition to Photoexcited PbS Colloidal Quantum Dots. *J. Phys. Chem. C* **2014**, *118* (38), 22284–22290.

(7)    Atkins, P. W.; Friedman, R. S. *Molecular Quantum Mechanics*; OUP Oxford, 2011.

(8)    Jadczak, J.; Kutrowska-Girzycka, J.; Smoleński, T.; Kossacki, P.; Huang, Y. S.; Bryja, L. Exciton Binding Energy and Hydrogenic Rydberg Series in Layered ReS2. *Sci. Rep.* **2019**, *9* (1).

(9)     Liu, E.; Long, M.; Zeng, J.; Luo, W.; Wang, Y.; Pan, Y.; Zhou, W.; Wang, B.; Hu, W.; Ni, Z.;

        You, Y.; Zhang, X.; Qin, S.; Shi, Y.; Watanabe, K.; Taniguchi, T.; Yuan, H.; Hwang, H. Y.;

        Cui, Y.; Miao, F.; Xing, D. High Responsivity Phototransistors Based on Few-Layer ReS2

        for Weak Signal Detection. *Adv. Funct. Mater.* **2016**, *26* (12), 1938–1944.

(10)    Sim, S.; Shin, H.-S.; Lee, D.; Lee, J.; Cha, M.; Lee, K.; Choi, H. Opposite Behavior of

        Ultrafast Dynamics of Exciton Shift and Linewidth Broadening in Bilayer Re S 2. *Phys.*

        *Rev. B* **2021**, *103* (1).